\documentclass[letterpaper, 10pt, twocolumn, twoside]{article}
\usepackage[hmargin=0.7in,vmargin=1in]{geometry}

\usepackage[utf8]{inputenc}
\usepackage[T1]{fontenc}
\usepackage{titling}
\usepackage[affil-it]{authblk}

\usepackage[backend=biber,sorting=none]{biblatex}
\usepackage{float}
\usepackage{graphicx}
\usepackage{setspace}
\usepackage{amsmath}

\title{Amorphous metal oxide mixtures for high-$Q$ integrated nonlinear photonics}

\author[1]{Alexa R. Carollo \thanks{alexa.carollo@nist.gov}}
\author[1,2]{Atasi Dan}
\author[1,2]{Haixin Liu}
\author[3]{David R. Carlson}
\author[1]{\authorcr Jennifer A. Black \thanks{Present address: Vescent Photonics, Golden, CO}}
\author[1,2]{Scott B. Papp}
\affil[1]{Time and Frequency Division, National Institute of Standards and Technology, Boulder, CO 80305 USA}
\affil[2]{Department of Physics, University of Colorado, Boulder, CO 80309, USA}
\affil[3]{Octave Photonics, Louisville, CO 80027, USA}

\date{}

\pretitle{\begin{center} \large \bfseries}
\posttitle{\par\end{center}}
\preauthor{\begin{center} \large}
\postauthor{\end{center}}

\begin{document}
\maketitle

\singlespacing 
\noindent\textbf{Choice of material is ubiquitous in integrated photonics to design device properties, whereas changing material composition is much less common. With phase matching as an additional objective, constraints in depositing and patterning thin films limit the use of integrated nonlinear photonics to only select materials. Here, we explore an amorphous metal oxide mixture of titania (TiO\textsubscript{2}) and tantala (Ta\textsubscript{2}O\textsubscript{5}) in which material composition is a tool to enhance and customize photonics properties. In particular, the inclusion of titania reduces the oxygen-defect density in a tantala film, while maintaining a comparable Kerr nonlinear index. With ion-beam sputtering at room temperature, we deposit a thick, ultralow-loss titania-tantala film, and we nanopattern it to create microresonator frequency combs. Titania-tantala microresonators offer lower loss, higher index of refraction, and reduced optical absorption and photorefractive effects.  Specifically, the titania-tantala mixture enables microresonator quality factor up to $10^7$ and a direct factor of 1.7 reduction in optical absorption. Our work demonstrates that composition in metal-oxide mixtures is a design parameter alongside the nanofabrication process and photonics design for integrated nonlinear photonics.} \vspace{10pt}

Integrated photonics offers various high-value functions, including dense and versatile circuits \cite{Molesky_2018}, access to modulation \cite{Boes_2023, Xu_2005} and photodetection \cite{Novack:13}, integration with microelectronics \cite{Rakowski:20}, and laser gain with heterogeneous integration \cite{Nader:25}. To take advantage of the unprecedented potential of integrated-photonics systems, low optical loss is critical for distributing light among complex circuits and between interfaces, placing constraints on photonic-material compatibility. Nonlinearity is an emerging function in integrated photonics that enables conversion from one wavelength to another and generation of quantum-based light states. Low loss and small mode volume are key to access nonlinearity, subject to control of group-velocity dispersion. The development of integrated nonlinear photonic microsystems drives innovation in low-loss materials, especially in improving their compatibility with integration and device geometric constraints, to open up opportunities in extreme-capacity data transmission \cite{Yang2022, Shirpurkar:22, Rizzo2023, Corcoran2025, Liu2024, Pirmoradi2025}, advanced computing architectures \cite{Xu2021, Feldmann2021}, quantum sensing \cite{Carlson:2017, Hsu:2022, Isichenko2023, Ferdinand:2025, Jammi:24, Niffenegger2020, Loh2020}, and signal generation \cite{Zang:2025, Kudelin2024, Sun2024}. Numerous discrete nonlinear photonics materials have been explored, and realizing fully functional platforms so far requires the use of different materials to overcome limitations in, for example, transparency or processing requirements.

To discern the tradeoffs between fabrication compatibility of materials and low optical loss, we survey the properties of different material platforms. Stoichiometric silicon nitride (Si$_3$N$_4$, or SiN) is an exemplary material for integrated nonlinear photonics, owing to relatively high refractive index for device design and group-velocity dispersion engineering, relatively high nonlinear coefficient, access to microresonator intrinsic quality factor, $Q_i$, above 10 million, and high power handling without photorefraction and multi-photon absorption \cite{Ye:19, Liu:25, Wilson2009, Ji2021, Liu:22, Corato-Zanarella:24, Ji:24}. However, the end-to-end process with low-pressure chemical-vapor-deposited (LPCVD) SiN requires high temperature 1200 \textcelsius{} annealing and the films have substantial tensile stress, compromising integration compatibility with other materials. Other formulations and deposition techniques, e.g. sputtering \cite{Zhang:2024} or plasma-enhanced CVD \cite{Bose:24}, lower the temperature requirement and enhance integration compatibility, but result in higher absorption that degrades microresonator $Q_i$. Crystalline materials like lithium niobate (LN) and semiconductors like gallium arsenide (GaAs) offer a wide variety of functionalities, but require complex wafer bonding and processing temperature constraints. The trend of such tradeoffs is common in integrated photonics, especially in gaining access to nonlinearity.

\begin{figure*}[t!]
\centering
    \includegraphics[scale=0.29]{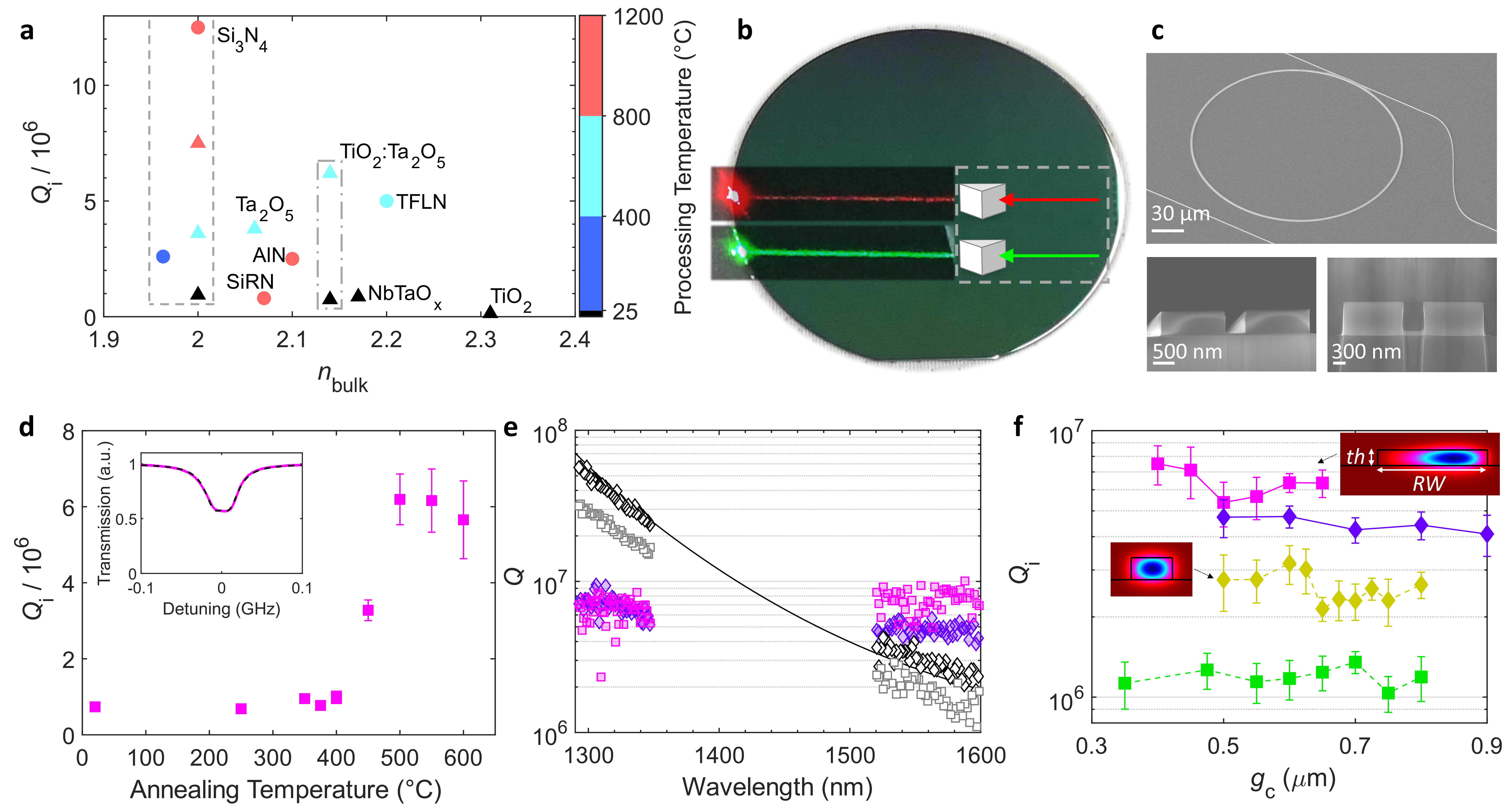}
    \caption{Characterization of titania-tantala films and microresonators. (a) Intrinsic $Q$ and $n$\textsubscript{bulk} at 1550 nm, comparing different photonics materials with triangles for sputtered films and point color for processing temperature. Repeated points indicate the sensitivity of $Q_i$ to fabrication and processing conditions. Increased $Q_i$ is reported for Si\textsubscript{3}N\textsubscript{4} with non-standard fabrication approaches \cite{Ji:17, LiuHuang:2021}, which we indicate by the open box. We label titania-tantala as TiO\textsubscript{2}:Ta\textsubscript{2}O\textsubscript{5}, silicon-rich silicon nitride as SiRN, thin-film lithium niobate as TFLN, and niobium-tantalum oxide as NbTaO\textsubscript{x}. (b) An oxidized silicon wafer coated with a titania-tantala film. The overlays illustrate prism coupling of laser light into the film at wavelengths of 637 nm (top) and 520 nm (bottom). (c) Scanning-electron microscope (SEM) image of a microresonator (top), and cross sections of two waveguides separated by 0.500 \textmu m with air (bottom left) and oxide cladding (bottom right). (d) $Q_i$ as a function of post-fabrication annealing temperature. Inset: mode transmission spectrum (pink) and its fit (black). (e) $Q_i$ (colors) and $Q_c$ (grayscale) as a function of wavelength for devices with different claddings. Squares indicate air cladding and diamonds indicate oxide cladding. The black line is an FDTD simulation of $Q_c$, scaled by 0.6, for the oxide clad device. (f) $Q_i$ as a function of $g_c$ for air clad (squares) and oxide clad (diamonds) devices with different $RW$. Points connected by solid and dashed lines are $RW$ = 4 \textmu m and $RW$ = 1.5 \textmu m, respectively. Insets: simulated electric field amplitude.}
    \label{fig:Paper1_Fig1}
\end{figure*}

While adjusting material composition is not common with ultralow-loss SiN photonics, semiconductor materials like GaAs and crystalline materials like LN are often customized with dopants. GaAs and its alloys make up an exceptional platform for laser gain, both in discrete epitaxial structures and through heterogeneous integration of epitaxial waveguide layers to low-loss passive and nonlinear materials \cite{Shekhar2024}. Furthermore, doping LN with magnesium oxide offers lower optical losses in the visible and reduced photorefractive effects for improved stability of its optical properties over undoped LN \cite{Zelmon:97}. On the other hand, amorphous metal oxide mixtures form a class of low-loss dielectric materials that offers an exceptional range of variation in composition \cite{Fazio:22}. There are numerous metal oxides with unique optical properties, including tantala (Ta\textsubscript{2}O\textsubscript{5}) \cite{Black:21, Lamee:20, Jung:21}, germanium oxide \cite{Yang:2021}, tungsten trioxide \cite{XUE2015127}, and barium titanate \cite{Chelladurai:2025}. Compared to SiN, tantala offers a wider transparency window \cite{Spektor:23}, a larger Kerr coefficient, and a lower processing temperature requirement. With metal oxides, sputtering deposition at low temperature and on arbitrary substrates is common \cite{MacFarlane:22, Irvine:25}.

Here, we show that material composition in metal-oxide mixtures is a viable design parameter for optimizing integrated nonlinear photonics. Specifically, we explore an amorphous film that is a mixture of two metal oxides, titania (TiO\textsubscript{2}) and tantala, motivated by use of this mixture in LIGO mirrors due to lower mechanical loss \cite{Pinard:17}. Titania-tantala features an intrinsic reduction in oxygen-vacancy defects, resulting in reduced optical absorption and photorefractive effects while maintaining a high nonlinear index. We demonstrate a robust, wafer-scale, CMOS-compatible nanofabrication process flow with all steps operated at $< 500$ \textcelsius{} that realizes microresonators with $Q_i$ up to $1 \times 10^7$. Quality factors of this magnitude are not readily achievable in other materials, including pure tantala, with comparable processing temperatures. Our results introduce a new method for controlling material properties, and we demonstrate progress towards overcoming the high-temperature requirements of ultralow-loss platforms. This work has the potential to enable fully integrated on-chip systems for future applications in optical frequency metrology, precision spectroscopy, and quantum sensing.

The principal advantage of titania-tantala for integrated nonlinear photonics is the combination of moderately high index of refraction, a modest process temperature requirement, and record high quality factor. In this work, we begin with titania-tantala films deposited on 3 inch oxidized silicon wafers via ion-beam sputtering with a 0.26 volume fraction of titania. We put the properties of titania-tantala into context by plotting microresonator $Q_i$ versus bulk index of refraction, $n$\textsubscript{bulk}, for several comparable materials in the 1550 nm wavelength region; see Fig. \ref{fig:Paper1_Fig1}a \cite{Ye:19, Zhang:2024, Bose:24, LiuHuang:2021, Jung:21, MacFarlane:22, YeFu:19, Liu:17, Shams-Ansari:2022, Fu:20, Guan:18}. Deposition technique and processing temperature are the factors that control integration of nonlinear materials. To date, silicon nitride provides the highest $Q_i$, but only when annealed at 1200 \textcelsius{}. Titania-tantala is promising because the temperature requirement to achieve high $Q_i$ is much more favorable. Even compared to pure tantala, titania-tantala offers the highest $Q_i$ of materials processed at moderate temperatures ($\leq 500$ \textcelsius{}). The effect of annealing metal oxide mixtures is primarily a reduction in absorptive oxygen-vacancy defects. Apparently, this is obtained at a much lower temperature than that required to remove hydrogen impurities in silicon nitride. Even without annealing, the intrinsic defect density of titania-tantala is low enough to provide a $Q_i$ of $7\times10^5$ with full room temperature processing. Moreover, un-annealed titania-tantala offers low visible-band absorption. We couple visible laser light into the thin film with a prism and quantify propagation loss; see Fig. \ref{fig:Paper1_Fig1}b. Our measurements approach the instrument limit, achieving propagation loss $<0.2$ dB/cm. This flexibility in processing temperature and ability to fabricate materials with custom composition are unique advantages of sputtered metal oxide films.

To explore integrated photonics devices in titania-tantala films, we fabricate microresonators with a typical CMOS process, including electro-beam lithography, reactive ion etching, silicon dioxide (SiO\textsubscript{2}, hereafter oxide) cladding deposition, and thermal annealing. This yields microresonators and integrated waveguide couplers; see Fig. \ref{fig:Paper1_Fig1}c.  The primary consideration in designing microresonators is for nonlinear optics, so we choose the titania-tantala film thickness $th=0.570$ \textmu m for air clad devices and $th=0.800$ \textmu m for oxide clad devices. These thicknesses enable fine-tuning of the group-velocity dispersion (GVD) throughout the anomalous and normal regimes by changing the ring width ($RW$) and ring radius ($RR$) of resonators. For external coupling of microresonators, we use pulley couplers that are optimized for broadband and selective coupling of the TE0 mode \cite{Zang:24}. To assess the etch quality, we fabricate test structures and take SEM images of their cross sections. In both air and oxide clad devices, we observe slight deviations from the designed rectangular profiles due to low etch selectivity.

Since annealing is critical to increase $Q_i$ in titania-tantala resonators, we characterize its temperature dependence, while maintaining appropriate annealing time and gas composition; see Fig. \ref{fig:Paper1_Fig1}d. In metal oxide films, oxygen-vacancy defects form during film deposition, exposure to vacuum, and exposure to UV radiation. The defects are absorptive and reduce $Q_i$ below the true material limit. Their structure and distribution amongst the surface and bulk is widely studied to control and optimize electrical and optical properties. For example, a common defect in amorphous films is TaO\textsubscript{2}, which is oxygen deficient compared to tantala (Ta\textsubscript{2}O\textsubscript{5}) \cite{Pedersen:2020}. As a way to increase the oxygen content, the films are annealed in air. This drives diffusion and chemical reactions with oxygen that reduce the defect density. We explore the effects of annealing in air clad microresonators with $RW=4\,\mu$m and $RR=100\,\mu$m. Indeed, our experiments reveal temperature-dependent behavior of $Q_i$ that reflects changes in material absorption.  We find a rapid increase in $Q_i$ up to $(3.3 \pm 0.3)\times10^6$ by annealing at 450 \textcelsius{}. This sharp transition indicates that oxygen has ample energy to diffuse into the film and react with oxygen-vacancy defects. The effect of annealing saturates above 500 \textcelsius{}, with an average $Q_i = (6.2 \pm 0.7)\times10^6$. Our results show that titania-tantala has the flexibility to be used across a wide range of temperatures, with maximum $Q_i$ approaching the material absorption limit when annealed up to 600 \textcelsius{}. We anticipate crystallization of the film beyond 600-700 \textcelsius{} \cite{Fazio:20}, which has not been explored with the production of tantala photonics because of the likelihood of cracking.

\begin{figure*}[h]
\centering
    \includegraphics[scale=0.29]{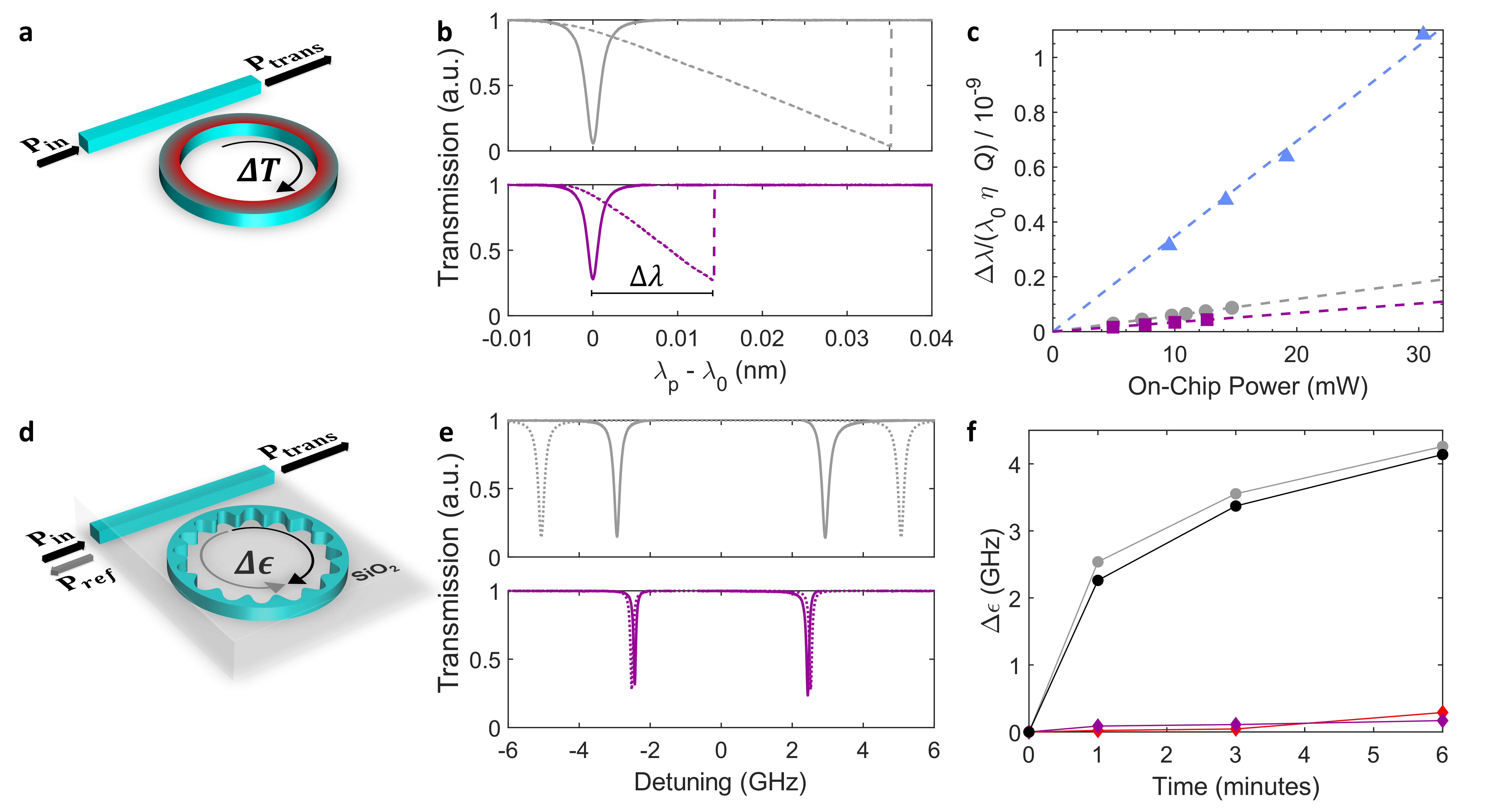}
    \caption{Exploring defect density in titania-tantala microresonators. (a) Microresonator with power-dependent absorption. $P_{\rm{in}}$: input power, $P_{\rm{trans}}$: transmitted power, $\Delta T$: change in temperature. (b) Power-dependent transmission lineshapes for tantala (gray, top plot) and titania-tantala (magenta, bottom plot) as a function of pump laser detuning, $\lambda_p - \lambda_0$. Solid lines: low power measurement. Dashed lines: high power measurement with $\approx10$ mW of on-chip power. The resonance shift due to heating, $\Delta\lambda$, is indicated on the bottom plot. (c) $\Delta\lambda$ (scaled) as a function of on-chip power for annealed tantala (gray circles), annealed titania-tantala (magenta squares), and unannealed titania-tantala (blue triangles) along with linear fits (dashed lines). (d) Illustration of an oxide clad PhCR. $P_{\rm{in}}$: input power, $P_{\rm{trans}}$: transmitted power, $P_{\rm{ref}}$: reflected power, $\Delta\epsilon$: change in mode splitting. (e) Measured transmission spectrum of a PhCR before pumping (solid lines). Dashed lines represent the change in transmission spectrum after pumping with $100$ mW of on-chip power for 6 minutes. Top plot: oxide clad tantala, bottom plot: oxide clad titania-tantala. (f) $\Delta \epsilon$ after pumping with $100$ mW on-chip power as a function of pumping time for two different tantala (gray and black circles) and titania-tantala (magenta and red diamonds) PhCR devices.}
    \label{fig:Paper1_Fig2}
\end{figure*}

With our optimized annealing process, we explore $Q_i$ in titania-tantala microresonators versus wavelength and the choice of either air or oxide cladding; see Fig. \ref{fig:Paper1_Fig1}e. Here, a number of factors influence $Q_i$, predominantly absorption in the titania-tantala and oxide cladding materials. The large $RW$ should substantially reduce scattering from the etched surfaces for both the 1300 nm and 1550 nm wavelength bands. The external coupling rate of the devices shifts between the two bands; hence, we plot the corresponding coupling quality factor, $Q_c$, of the $Q_i$ measurements as open points. An FDTD (finite-difference time-domain) calculation of $Q_c$ with cladding explains the measured results within a scale factor of 0.6. In the 1300 nm band, we expect the oxide cladding to have low absorption \cite{Bauters:11}. Comparing air and oxide clad $Q_i$ in this wavelength region allows us to assess potential damage to the titania-tantala layer induced by the oxide deposition process. We do not observe a significant difference; in fact, we achieve high $Q_i$ up to $9.5\times10^6$ in the oxide clad device. Hence, the film's robustness and our annealing process can ensure low loss, even after the oxide deposition step. In the 1550 nm band, where the mode is larger and extends further outside the material, the results are different. We achieve $Q_i$ up to $1\times10^7$ in the air clad device, but the $Q_i$ for the oxide clad resonators is lower. Absorption in the oxide cladding is the likely cause.

To further investigate the factors that limit $Q_i$, we systematically vary the microresonator device geometry to influence the relative impact of absorption and scattering losses; see Fig. \ref{fig:Paper1_Fig1}f. We have the capability to widely vary the external coupling rate through changes in the resonator-waveguide gap distance, $g_c$. This is an important factor in diagnosing loss mechanisms because it controls $Q_c$. In our largest $RW$ air clad microresonators, we vary $g_c$ from 0.4 \textmu m to 0.65 \textmu m. This induces a change in $Q_c$ from $2\times10^6$ to $6\times10^7$, while maintaining a constant $Q_i$ near $6\times10^6$. The $Q_i$ is unchanged over this range of $g_c$ for several settings of $RW$, demonstrating control of the coupling rate without introducing parasitic coupling loss. As we decrease the cross-sectional dimension of the ring, we reveal a transition from absorption-limited to scattering-limited $Q_i$. Decreasing the $RW$ to 1.5 \textmu m causes the mode profile of the propagating light to extend further into the sidewalls, which have rough surfaces from the etch process. The rough surfaces induce scattering, which degrades $Q_i$ below the absorption limit.

The central advantage of changing composition in a metal oxide mixture is to enhance device properties without otherwise modifying the fabrication process. Here, we directly test for reductions in optical absorption and effects like photorefraction that depend on defect density. For optical absorption, we measure the quality factor associated with absorption, $Q_{\rm{abs}}$, by microresonator thermal bistability. We directly compare titania-tantala and pure tantala microresonators to characterize their difference in defect density. Our thermal bistability measurements follow the canonical model of Ref. \cite{Carmon:04}. A microresonator mode resonance wavelength ($\lambda_0$) shifts with a change in temperature ($\Delta T$) as
\begin{equation}
    \lambda_r(\Delta T) = \lambda_0(1 + a\Delta T),
    \label{eq:lam_r}
\end{equation}
where $a = \alpha + (dn/dT)/n_0$ is the temperature coefficient, containing the thermal expansion coefficient ($\alpha$), the thermo-optic coefficient ($dn/dT$), and index of refraction ($n_0$). In our experiments, $\Delta T$ is induced by absorption of the pump laser in the material, as illustrated in Fig. \ref{fig:Paper1_Fig2}a. As the pump laser wavelength ($\lambda_p$) sweeps from lower to higher wavelength, the material dynamically heats and the mode resonance thermally broadens; see Fig. \ref{fig:Paper1_Fig2}b. Notably, the shift in resonance wavelength ($\Delta\lambda = \lambda_r - \lambda_0$) is larger in tantala than in titania-tantala, which indicates higher absorption and defect density in tantala.

We systematically measure $\Delta\lambda$ at various pump powers to extract $Q$\textsubscript{abs} for three different microresonators; see Fig. \ref{fig:Paper1_Fig2}c. Our primary goal is to compare absorption in titania-tantala and tantala, with a focus on annealed devices for highest $Q_i$. By also comparing with an un-annealed titania-tantala microresonator, we quantify the underlying absorption changes induced by annealing. At steady state and when $\lambda_p$ is on resonance with $\lambda_r$ (at the transmission minimum), $\Delta T = I\eta Q/(Q_{\rm{abs}}k)$, where $I$ is the on-chip pump power, $\eta$ is the coupling efficiency ($\eta = (1+1/K)^{-1}$), $Q$ is the total quality factor, and $k$ is the thermal conductivity of the mode volume with the surroundings. The coupling parameter, $K$, is defined as $K \equiv Q_i/Q_c$. This relationship, and the relationship between $\Delta T$ and $\Delta \lambda$ in equation (\ref{eq:lam_r}), allows us to directly extract information about $Q_{\rm{abs}}$. With corrections for differences in microresonators, we observe a linear dependence of our scaled $\Delta\lambda$ with $I$, with annealed titania-tantala yielding the smallest slope. The slope encompasses the remaining unknown parameters, $a/(Q_{\rm{abs}}k)$. To simplify our analysis, we assume that the thermal parameters, $a$ and $k$, are the same for all three films. With these assumptions, we can directly compare $Q_{\rm{abs}}$. As expected from the drastic improvement in titania-tantala's $Q_i$ with annealing, we find that $Q_{\rm{abs}}$ increases by a factor of 10 with annealing. This is consistent with the $>8$ times improvement in $Q_i$, validating our assumptions in this experiment. Compared to annealed tantala, annealed titania-tantala has a 1.7 times higher $Q_{\rm{abs}}$. This is also consistent with our $Q_i$ measurements, but this experiment provides direct evidence that metal oxide mixtures support intrinsically low defect density.

Photorefraction, an effect in which the material index of refraction changes with optical intensity, is a consequence of high defect density. In tantala microresonators, it occurs due to charge redistribution and build-up in defect sites. The effect induces coherent backscattering and optical interference of a large enough magnitude that can hinder GVD phase matching and reduce efficiency. Coherent backscattering causes a transient split of the pump resonance. The magnitude of the induced splitting depends on pumping time, $t$, as
\begin{equation}
    \epsilon (t) = \epsilon_0 + \xi(1-e^{-t/\tau_p})|\beta^*_+ \beta_-|
    \label{eq:pr}
\end{equation}
where $\epsilon_0$ is the static splitting, $\xi$ is a material parameter, $\tau_p$ is the time constant, and $|\beta^*_+ \beta_-|$ is the intensity of the standing wave generated by the forward ($\beta_+$) and backward ($\beta_-$) propagating fields in the resonator \cite{Liu:2021}. We assume that $\xi$, which controls the magnitude of the time-dependent part, is related to defect density in metal oxides. A higher defect density should have a larger photorefractive effect because there are more defect sites available for charges to accumulate.

In a photonic-crystal microresonator (PhCR), we intentionally inscribe a modulation of the $RW$ during device fabrication; see Fig. \ref{fig:Paper1_Fig2}d. The modulation induces coherent backscattering and generates a band gap with splitting of $\epsilon_0$, which facilitates soliton microcomb generation \cite{Yu:2021}. However, the standing wave created within the resonator is enhanced by the photorefractive effect, increasing the mode splitting by $\Delta\epsilon = \epsilon - \epsilon_0$ from the designed value. Oxide-clad tantala devices, in particular, suffer from significant photorefractive splitting; see Fig. \ref{fig:Paper1_Fig2}e. After pumping the PhCR mode with $I$ = 100 mW for 6 minutes, the $\Delta\epsilon$ is 4.2 GHz. On the contrary, oxide-clad titania-tantala is much more robust. The $\Delta\epsilon$ is only 0.17 GHz, which is 25 times less than tantala. To understand the dynamics, we investigate $\Delta\epsilon$ as a function of pumping time; see Fig. \ref{fig:Paper1_Fig2}f. For both tantala PhCRs (gray and black traces), $\Delta\epsilon$ exceeds 2 GHz after pumping for only 1 minute, and approaches saturation by 6 minutes. For both titania-tantala PhCRs (red and magenta traces), $\Delta\epsilon$ remains below 0.3 GHz at all times. The saturation dynamics are less apparent, perhaps due to the small $\Delta\epsilon$. Since the photorefractive splitting is significantly smaller in titania-tantala, this indicates that its defect density is intrinsically lower than tantala.

\begin{figure*}[t]
\centering
    \includegraphics[scale=0.29]{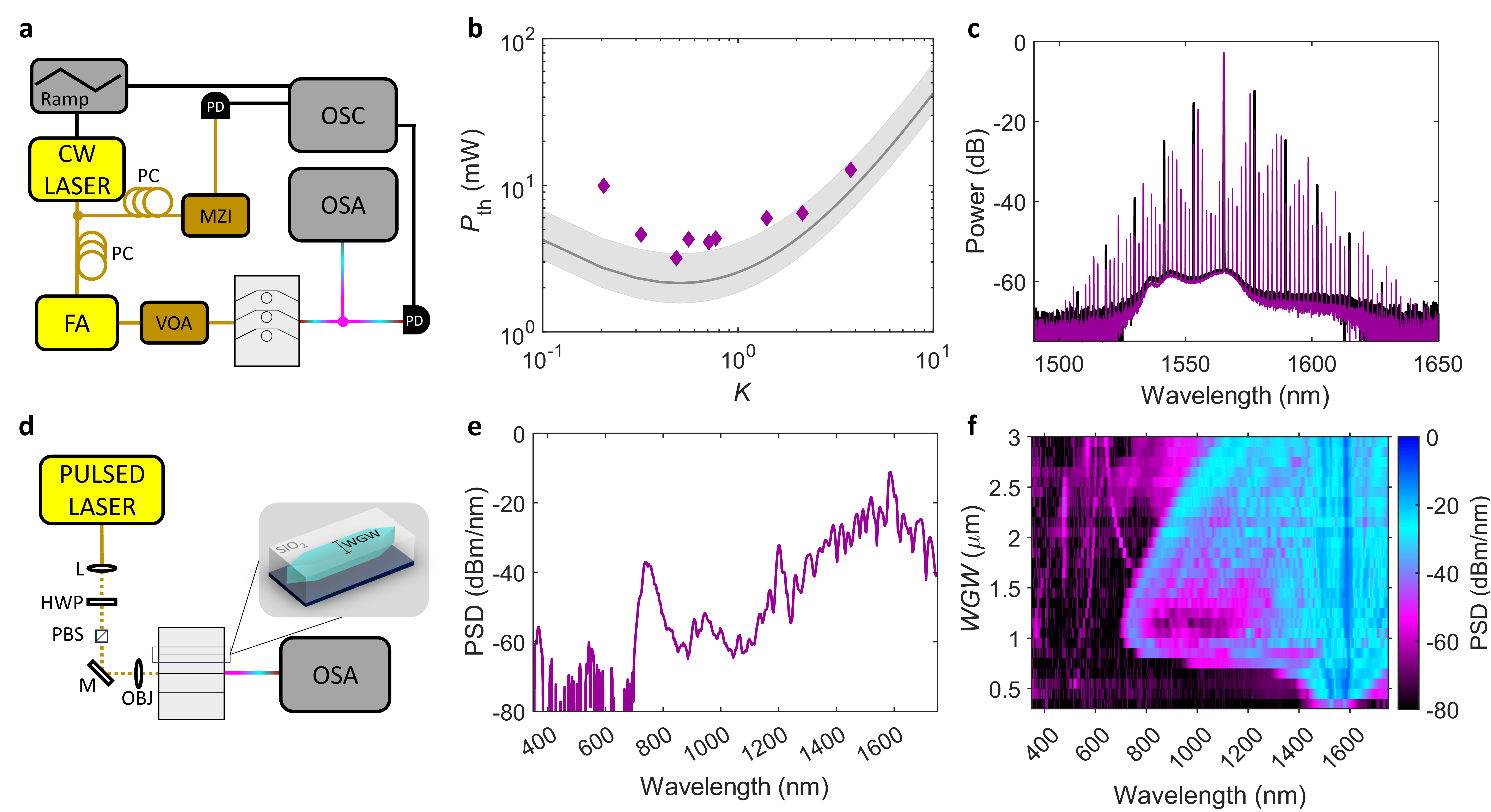}
    \caption{Nonlinearity in titania-tantala. (a) Experimental setup for generating OPO and microcombs. CW: continuous wave, PC: polarization control, FA: fiber amplifier, MZI: Mach-Zehnder interferometer, PD: photodiode, VOA: variable optical attenuator, OSC: oscilloscope, OSA: optical spectrum analyzer. (b) $P_{\rm{th}}$ as a function of $K$ for oxide clad microresonators (magenta diamonds), along with a calculation using $n_2$ of tantala (gray band). (c) Microcomb spectra from pumping a microresonator with 11 mW (black) and 17 mW (magenta) on-chip power, scaled relative to pump power (0 dB). (d) Experimental setup for generating supercontinuum. L: lens, HWP: half-waveplate, PBS: polarizing beamsplitter, M: mirror, OBJ: objective lens. (e) Supercontinuum spectrum generated from pumping a $WGW$ = 1 \textmu m oxide clad device with 0.6 nJ/pulse. (f) Contour plot of supercontinuum spectra as a function of $WGW$.}
    \label{fig:Paper1_Fig3}
\end{figure*}

Our characterization of the titania-tantala metal oxide mixture indicates that composition does offer the opportunity to enhance and customize integrated photonics properties. Here, we measure the threshold and operation of nonlinear wavelength conversion in titania-tantala to assess the Kerr nonlinear index of the mixture in comparison with pure tantala. To measure the Kerr nonlinear index, we explore the threshold power for microresonator optical parametric oscillation (OPO) and the generation of supercontinuum from an input modelocked laser.

To generate OPO in titania-tantala microresonators, we select devices of a fixed resonator geometry, and we use variation in the external coupling to systemically study the OPO threshold power. To satisfy the phase-matching conditions for OPO generation, we design the geometric parameters to support anomalous GVD. For a microresonator with oxide cladding, we achieve this with a $RR=100$ \textmu m and $RW=1.5$ \textmu m. We pump the microresonator with a continuous-wave (CW) laser, using the experimental arrangement in Fig. \ref{fig:Paper1_Fig3}a, and we record the pump power for OPO initiation. The OPO threshold power depends on the Kerr nonlinear index, $n_2$, as
\begin{equation}
    P_{\rm{th}} = \frac{\pi n_0 \omega_0 A_{\rm{eff}}}{4 n_2 D_1 Q_i^2} \frac{(K+1)^3}{K}
    \label{eq:pth}
\end{equation}
where $\omega_0$ is the optical frequency, $A_{\rm{eff}}$ is the effective mode area, and $D_1$ is the free spectral range. Indeed, the geometry of the microresonator determines many of the parameters that influence $P_{\rm{th}}$. For a fixed geometry, a way to optimize $P_{\rm{th}}$ is through adjusting the external coupling. We vary $g_c$ to adjust the coupling parameter, $K \equiv Q_i/Q_c$, and measure its effect on $P_{\rm{th}}$; see Fig. \ref{fig:Paper1_Fig3}b. As predicted from equation (\ref{eq:pth}), we observe a dependence on $K$, with a minimum at $K=0.5$. Our prediction uses the $n_2$ of tantala, which is $(6.2\pm 2.3)\times10^{-19}$ m\textsuperscript{2}/W \cite{Jung:21}. Since our measured values fall near the predicted values across a range of $K$, we estimate that the $n_2$ of titania-tantala is comparable to tantala. Apparently, the modification to the film composition has a minimal effect on $n_2$. When pumping above the threshold, we observe the onset of microcomb formation; see Fig. \ref{fig:Paper1_Fig3}c.

To further explore the Kerr nonlinearity of the titania-tantala mixture, we generate supercontinuum in a  waveguide and characterize its spectral properties. The high confinement in waveguides naturally allows access to nonlinear processes that broaden the spectrum of an input modelocked laser. We use a 1550 nm band modelocked laser to pump a fully oxide clad titania-tantala waveguide and record its output spectrum by use of the experimental setup in Fig. \ref{fig:Paper1_Fig3}d. We design the waveguide width ($WGW$) such that the GVD supports a broad supercontinuum; see Fig. \ref{fig:Paper1_Fig3}e. Notably, the power spectral density (PSD) in the 750 nm wavelength region is enhanced by a dispersive wave. We control the wavelength of the dispersive wave by changing the $WGW$, as demonstrated in Fig. \ref{fig:Paper1_Fig3}f. As we vary the $WGW$ from 1 \textmu m to 3 \textmu m, the dispersive wave shifts from 750 nm to 1100 nm. The input pulse energy required to generate these spectra is 0.6 nJ/pulse, which is comparable to tantala \cite{Lamee:20}. Since $n_2$ controls the energy required for supercontinuum generation, these results provide further evidence that the nonlinear behavior of the mixture is comparable to pure tantala.

In conclusion, we have demonstrated a powerful method to engineer material properties for integrated nonlinear photonics by tailoring the composition of amorphous metal oxide films. By introducing titania into tantala, we reduce defect density without altering the low-temperature, CMOS-compatible fabrication process. This composition-driven approach yields record-high quality factors at $<500$ \textcelsius{}, significantly lowering absorption and suppressing photorefractive effects, while preserving the high Kerr nonlinearity of tantala. These improvements position titania-tantala as an enabling material for low-loss, nonlinear, and dispersive photonic devices. Looking ahead, this work establishes a new pathway to scalable, monolithically integrated photonic systems for next-generation applications in frequency metrology, communications, and quantum technologies.

We acknowledge Zachary Newman and Cecile Carlson from Octave Photonics for supercontinuum measurements. We thank Sarang Yeola and Nima Nader for technical review of the letter. This research has been funded by the AFOSR FA9550-20-1-0004 Project Number 19RT1019, NSF Quantum Leap Challenge Institute Award OMA – 2016244, DARPA LUMOS (HR0011-20-2-0046) and NaPSAC, Microelectronics Commons NW-AI-Hub, and NIST. Trade names provide information, not an endorsement.

\printbibliography

\clearpage
\section{Methods}

\subsection{Material platform comparison}
To compare different materials within the context of integrated nonlinear photonics, we explore $Q_i$ versus $n_{\text{bulk}}$ and note the required processing temperature. These factors bring together considerations in integrating different photonics materials and achieving phase matching for four-wave mixing. 
We plot average $Q_i$ if reported; otherwise, we plot highest $Q_i$ in Fig. \ref{fig:Paper1_Fig1}a. We define processing temperature as the highest of reported annealing or deposition temperature.

\subsection{Fabrication}
Taking a wafer as in Fig. \ref{fig:Paper1_Fig1}b, our fabrication process flow includes electron-beam lithography to create a pattern and deposition of alumina to create a hard mask, dry reactive ion etching (RIE) with fluorine, deep RIE for chip separation, and in some cases, annealing for 10 hours at 500 \textcelsius{} in air. Our process is amenable to SiO\textsubscript{2} cladding deposition with an inductively coupled plasma CVD process that offers uniformity, low loss, and low void formation.

\subsection{Quality factor measurement}
We measure microresonator $Q_i$ for the TE0 mode family, identified by polarization and free-spectral range, over the wavelength range of 1520-1600 nm to sample many modes. We measure transmission across this range and fit each resonance lineshape to a model to extract $Q_i$ \cite{Gorodetsky:00}, as shown in the inset in Fig. \ref{fig:Paper1_Fig1}d. Here, we set $Q_c$ such that the coupling parameter $K \approx 0.25 - 0.45$ for a consistent measurement across devices with different annealing conditions.

\subsection{Absorption measurement}
To correct for differences in the microresonators, we scale $\Delta\lambda$ by known experimental parameters $\lambda_0$, $\eta$, and $Q$ in Fig. \ref{fig:Paper1_Fig2}c. We also correct for facet loss (typically 3-5 dB/facet) for a more accurate estimate of $I$. We assume that $a$ and $k$ are the same for all films because they are sputtered by the same process, the microresonators have similar geometries ($th$ = 0.570 \textmu m, $RR$ = 100 \textmu m, $RW$ = 1.5 or 1.6 \textmu m), and they have similar thermo-optic coefficients \cite{Jung:21} \cite{Carollo:23}.

\subsection{Photorefraction measurement}
In Fig. \ref{fig:Paper1_Fig2}e, we measure the steady-state band gap transmission while sweeping a tunable laser across the resonance (solid traces). We fit the band gap mode to extract $\epsilon_0$. We monitor changes to the mode structure that are induced by pumping the PhCR mode with $I$ = 100 mW on an oscilloscope. To represent the changes after $t=6$ minutes, we incorporate $\Delta\epsilon$ into the fit (dashed lines).

\end{document}